\documentclass{iopconfser}

\usepackage{array}
\usepackage{latexsym,graphicx,epsfig,psfrag}
\usepackage{amssymb,amsmath,amsxtra,amsfonts}

\newcommand{\Tr}{\mbox{\rm{tr}}}
\newcommand{\slp}{p\kern-5pt/}
\newcommand{\nn}{\nonumber\\}

\begin{document}

\title{Polarization observables in semileptonic $V\to P$ decays of quarkonia}

\author{C.~T.~Tran$^{1}$, M.~A.~Ivanov$^{2}$, N.~D. Nguyen$^{3}$ and Q.~C.~Vo$^{4}$}

\affil{$^1$Department of Physics, HCMC University of Technology and Education, Vo Van Ngan 1, 700000 Ho Chi Minh City, Vietnam}
\affil{$^2$Bogoliubov Laboratory of Theoretical Physics, Joint Institute for Nuclear Research, 141980 Dubna, Russia}
\affil{$^3$Department of Physics, HCMC University of Education, An Duong Vuong 280, 700000 Ho Chi Minh City, Vietnam}
\affil{$^4$University of Science, 700000 Ho Chi Minh City, Vietnam}

\email{thangtc@hcmute.edu.vn}

\begin{abstract}
We investigate the weak semileptonic decays of the charmonium $J/\psi$ and bottomonium $\Upsilon(1S)$ into a pseudoscalar meson, namely, the $J/\psi\to D_{(s)}\ell\bar{\nu}_\ell$ and $\Upsilon(1S)\to B_{(c)}\ell\bar{\nu}_\ell$ decays. We focus on the polarization observables including the forward-backward asymmetry, final lepton polarization, and lepton-side convexity parameter. In order to define these quantities, we apply the helicity amplitude method and obtain a twofold angular decay distribution for a general $V\to P\ell\bar{\nu}_\ell$ case. The hadronic form factors are taken from our previous work, where they were calculated in the Covariant Confined Quark Model. We provide theoretical predictions for the polarization observables and compare them with the literature.
\end{abstract}

\section{Introduction}
The charmonium $J/\psi$ and bottomonium $\Upsilon(1S)$ have large masses ($m_{J/\psi}=3096.90(10)~\textrm{MeV}$, $m_{\Upsilon(1S)}=9460.40(10)~\textrm{MeV}$) but narrow total decay widths ($\Gamma_{J/\psi}=92.6 (1.7)~\textrm{keV}$, $\Gamma_{\Upsilon(1S)}=54.02 (1.25)~\textrm{keV}$)~\cite{ParticleDataGroup:2024cfk}. They belong to the group of low-lying quarkonia. Low-lying quarkonia refer to the bound states of a heavy quark and its corresponding antiquark, such as charmonium (composed of a charm quark-antiquark pair) and bottomonium (composed of a bottom quark-antiquark pair). These are considered ``low-lying" because they are among the lowest energy states of these quark-antiquark systems. The discoveries of the charmonium $J/\psi$~\cite{E598:1974sol} and bottomonium $\Upsilon(1S)$\cite{SLAC-SP-017:1974ind} as heavy resonances with very narrow widths played an important role in the development of quantum chromodynamics (QCD) at early stages. These heavy quarkonia have been the object of an enormous amount of studies, both experimental and theoretical. One of the key features of heavy quarkonia that attract the theoretical community is the possibility to seperate the scales characterizing these bound states. These scales include the hard scale (related to the heavy quark mass), the soft scale (related to the small relative momentum of the quark-antiquark, $|\vec{p}|\sim mv$, $v\ll 1$), and the ultrasoft scale (related to the kinetic energy $E\sim mv^2$ of the heavy quark and antiquark). Another important feature of heavy quarkonia is that the quark and antiquark have masses much larger than the hadronic scale $m\gg \Lambda_{\textrm{QCD}}$. Therefore, the perturbation approach can be applied to study processes happening at the quark mass scale. It is also possible to use the scale hierachies $m\gg |\vec{p}|\gg E$ and $m\gg \Lambda_{\textrm{QCD}}$ to develop effective field theories (EFTs) to study low energy QCD in a systematic manner~\cite{Bodwin:1994jh}. A detailed review of EFTs for heavy quarkonia is given by Brambilla \emph{et al.} in Ref.~\cite{Brambilla:2004jw}.

Low-lying quarkonia mainly decay through intermediate photons or gluons produced by the parent $q\bar{q}$ quark pair annihilation. As a result, strong and electromagnetic modes dominate their decay process. For example, dominating decay modes of the bottomonium $\Upsilon(1S)$ include the hadronic decays via a three-gluon mediate state with a branching fraction of $\mathcal{B}(\Upsilon(1S)\to ggg) = (81.7\pm 0.7)$\%~\cite{CLEO:2005mdr,ParticleDataGroup:2024cfk}, the hadronic decays via two gluons plus a single photon  $\mathcal{B}(\Upsilon(1S)\to \gamma gg) = (2.2\pm 0.6)$\%, and the electromagnetic leptonic decays mediated by an offshell photon with branching fractions $\mathcal{B}(\Upsilon(1S)\to \tau^+\tau^-) = (2.60\pm 0.10)$\%, $\mathcal{B}(\Upsilon(1S)\to \mu^+\mu^-) = (2.48\pm 0.04)$\%, and $\mathcal{B}(\Upsilon(1S)\to e^+e^-) = (2.39\pm 0.08)$\%~\cite{ParticleDataGroup:2024cfk}. In contrast to strong and electromagnetic decays which have been studied extensively, weak decays of quarkonia have attracted less attention from the particle physics community since these processes are rare. However, in the last few years the BESIII has been intensively searched for experimental evidence of the rare weak semileptonic decays of the charmonium $J/\psi$ and obtained better limits for the branching fractions of $J/\psi\to D\ell^+\nu_{\ell}$ ($\ell = e,\mu$)~\cite{BESIII:2021mnd,BESIII:2023fqz}. Note that the up-comming Super Charm-Tau factory also includes the semileptonic decays of $J/\psi$ in its physical program~\cite{Achasov:2024eua}. This increases the expectation for experimental study of similar decays of $\Upsilon(1S)$ in the near future.

Semileptonic decays of the quarkonia $J/\psi$ and $\Upsilon(1S)$ have been studied in the framework of the Isgur-Scora-Grinstein-Wise (ISGW) model~\cite{SanchisLozano:1993ki, Kurdadze:1997ra}, Bauer-Stech-Wirbel (BSW) model~\cite{Kurdadze:1997ra,Dhir:2009rb}, light front quark model (LFQM)~\cite{ Shen:2008zzb, Chang:2019obq}, Bethe-Salpeter method~\cite{Wang:2016dkd}, nonrelativistic QCD (NRQCD)~\cite{Chang:2016gyw}, QCD sum rules (QCDSR)~\cite{Wang:2007ys}, lattice QCD (LQCD)~\cite{Meng:2024nyo},
and the covariant confined quark model (CCQM)~\cite{Ivanov:2015woa, Tran:2024phq} which had been developed previously by our group. Most studies (including ours) focused on the decay width since these decays are rare and the decay width is the most important observable. However, with the development of experimental facilities such as BESIII and Super Charm-Tau factory, more observables in these decays will soon be measured. Note that very few studies have been carried out for these polarization observables. To our knowledge, there is only one paper that addressed this question, which is Ref.~\cite{Chang:2016gyw}. It is therefore necessary to provide more theoretical predictions for polarization observables in these channels. They will be useful for future tests of the Standard Model (SM), as well as for searching for New Physics beyond the SM. In this study, we aim at the forward-backward asymmetry, final lepton polarization, and lepton-side convexity parameter. They are defined through a general twofold angular decay distribution obtained by using the helicity amplitude technique. We will provide prediction for their $q^2$ dependence as well as $q^2$-average value and compare with the literature. We note that, upon completing this manuscript, a very recent paper -- also considering polarization observables in the semileptonic decay $\Upsilon(1S)\to B_{(c)}\ell\bar{\nu}_\ell$ -- was published~\cite{Sheng:2025duk}. We will therefore compare our predictions with those in this reference when presenting our numerical results 

The crucial element in theoretical calculations of the semileptonic decays of quarkonia is the  invariant form factors describing the $V\to P$ hadronic transitions. These are functions (of momentum transfer between participating hadrons) that encapsulate the complex internal structure and dynamics of hadrons during their transitions and therefore determine the amplitude of the last. The calculation of hadronic form factors requires nonperturbative methods such as lattice QCD (LQCD), QCD sum rules (QCDSR), and phenomenological quark models. In this study, we use the form factors calculated in our previous studies~\cite{Ivanov:2015woa, Tran:2024phq} using the CCQM. The last is a quantum-field-theory--inspired phenomenological quark model built upon an interacting Lagrangian between a hadron and its constituent quarks (see, e.g., Refs.~\cite{Branz:2009cd, Ivanov:2011aa, Tran:2023hrn, Groote:2021ayy, Ivanov:2020iad}). 

The rest of the paper is structured as follows. In Sec.~\ref{sec:formalism} we present the relevant theoretical formalism for the calculation of the semileptonic decays $V\to P\ell\bar{\nu}_\ell$. The section includes the introduction of the helicity amplitude technique, the parametrization of the hadronic amplitude using form factors, and the derivation of the twofold differential decay distribution. In Sec.~\ref{sec:result} we define polarization observables based on the twofold decay distribution and provide their numerical results. A comparison with the literature is also provided. Finally, a brief summary is given in Sec.~\ref{sec:sum}.

\section{Helicity amplitudes and angular decay distribution}
\label{sec:formalism}
\begin{figure}[h]
	\centering
	\begin{tabular}{c} 
		\includegraphics[width=0.4\textwidth]{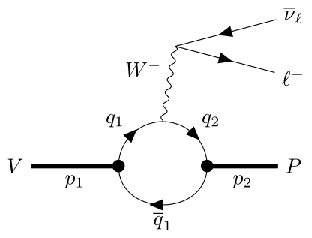}
	\end{tabular}
	\caption{\label{fig:semilept}
		Feynman diagram for semileptonic decays $V\to P\ell\bar{\nu}_\ell$.}
\end{figure}

The weak semileptonic decays $J/\psi\to D_{(s)}\ell\bar{\nu}_\ell$ and $\Upsilon(1S)\to B_{(c)}\ell\bar{\nu}_\ell$ are induced by the quark-level transitions $c \to d(s) \ell \bar{\nu}_{\ell}$ and $b \to u(c) \ell \bar{\nu}_{\ell}$ as described by the Feynman diagram in Fig.~\ref{fig:semilept} with the corresponding effective Hamiltonian
\begin{equation}
	\mathcal{H}_{\textrm{eff}} = 
	\frac{G_F}{\sqrt{2}} V_{q_1q_2} 
	\left[ \bar{q_2} \gamma_\mu(1-\gamma_5)  q_1 \right] 
	\left[ \bar{\ell} \gamma^\mu(1-\gamma_5)  \nu_{\ell} \right],
	\label{eq:Hamiltonian}
\end{equation}
where $q_1 = c\, \&\, q_2 = (d, s)$ for the $J/\psi\to D_{(s)}$, and $q_1 = b \, \&\, q_2 = (u, c)$ for the $\Upsilon(1S)\to B_{(c)}$ transitions, respectively, $G_F$ is the Fermi constant, and $V_{q_1q_2}$ is the corresponding Cabibbo-Kobayashi-Maskawa (CKM) matrix element.

The invariant matrix element of the decays is given by
\begin{equation}
	\mathcal{M}  =  
	\frac{G_F}{\sqrt{2}} V_{q_1q_2}
	\left\langle P|\bar{q_2}\gamma_\mu(1-\gamma_5) q_1|V\right\rangle
	\bar\ell \gamma^\mu(1-\gamma_5) \nu_\ell.
	\label{eq:M}
\end{equation}
\vspace*{-0.5cm}

The hadronic matrix element in Eq.~(\ref{eq:M}) is often parametrized as a linear combination of Lorentz structures multiplied by scalar functions, namely, invariant form factors which depend on the momentum transfer squared. For the $V \to P$  transition one has
\begin{align}
	&\left\langle P|\bar{q_2}\gamma_\mu(1-\gamma_5) q_1|V\right\rangle \equiv \epsilon^\nu_1 T^{VP}_{\mu\nu}
	\nn
	&=\frac{\epsilon_1^{\nu}}{m_1 + m_2}
	[ - g_{\mu\nu}pqA_0(q^2) + p_{\mu}p_{\nu}A_+(q^2)+q_{\mu}p_{\nu}
	A_-(q^2) + i \varepsilon_{\mu\nu\alpha\beta}p^{\alpha}q^{\beta}V(q^2)],
	\label{eq:FFpseudoscalar}
\end{align}
where $q=p_1-p_2$, $p=p_1+p_2$, $m_1 = m_V$, $m_2 = m_P$, and $\epsilon_1$ is the polarization vector of $V$, so that $\epsilon_1\cdot p_1 = 0$. The mesons are on their mass shells, i.e., $p_1^2=m_1^2=m^2_V$ and $p_2^2=m_2^2=m^2_P$. We use the following convention for the Levi-Civita tensor in Minkowski space: 
\begin{eqnarray}
	&&
	\Tr\left(\gamma_5\gamma^\mu\gamma^\nu\gamma^\alpha\gamma^\beta \right)
	= 4i\varepsilon^{\mu\nu\alpha\beta},
	\qquad
	\Tr\left(\gamma_5\gamma_\mu\gamma_\nu\gamma_\alpha\gamma_\beta \right)
	= 4i \varepsilon_{\mu\nu\alpha\beta},
	\qquad
	\varepsilon_{0123} = -\varepsilon^{0123} = +1.
\end{eqnarray}
The form factors $A_0(q^2)$, $A_{\pm}(q^2)$, and $V(q^2)$ for the decays $J/\psi\to D_{(s)}\ell\bar{\nu}_\ell$ and $\Upsilon(1S)\to B_{(c)}\ell\bar{\nu}_\ell$ being used in this paper were calculated in our previous studies~\cite{Ivanov:2015woa, Tran:2024phq} using the CCQM. The details of the model have been thoroughly discussed in numerous studies by our group. Here, we simply take the form factors from Refs.~\cite{Ivanov:2015woa, Tran:2024phq} to maintain the focus on the new findings.

The squared matrix element can be written as a product of the hadronic tensor $H_{\mu\nu}$ and leptonic tensor $L^{\mu\nu}$:
\begin{equation}
	\label{eq:M2}
	|\mathcal{M}|^2 = \frac{G_F^2 |V_{q_1q_2}|^2}{2}  H_{\mu\nu} L^{\mu\nu}.
\end{equation}
The leptonic tensor for the process
$W^-_{\rm off-shell}\to \ell^-\bar \nu_\ell$ is given by
\begin{eqnarray}
	L^{\mu\nu} &=& 
	\Tr\left[ (\slp_\ell + m_\ell) \gamma^\mu(1-\gamma_5) \slp_{\nu_\ell} \gamma^\nu(1-\gamma_5)\right]
	\nn
	&=&
	8 \left( 
	p_\ell^\mu p_{\nu_\ell}^\nu  + p_\ell^\nu p_{\nu_\ell}^\mu 
	- p_{\ell}\cdot p_{\nu_\ell}g^{\mu\nu}
	+  i \varepsilon^{\mu \nu \alpha \beta} p_{\ell\alpha} p_{\nu_\ell\beta}
	\right).
	\label{eq:lept_tensor}
\end{eqnarray}
The hadronic tensor in Eq.~(\ref{eq:M2}) reads
\begin{equation}
	H_{\mu\nu} = 
	T^{VP}_{\mu\alpha}\left(-g^{\alpha\alpha'}+\frac{p_1^\alpha p_1^{\alpha'}}{m_1^2}\right)
	T^{VP\dagger}_{\nu\alpha'}, 
\end{equation}
where
\begin{equation}
	T^{VP}_{\mu\alpha}=\frac{1}{m_1 + m_2}
	\left[- g_{\mu\alpha}pqA_0(q^2) + p_{\mu}p_{\alpha}A_+(q^2)+q_{\mu}p_{\alpha}
	A_-(q^2) + i \varepsilon_{\mu\alpha\gamma\delta}p^{\gamma}q^{\delta}V(q^2)\right] .
\end{equation}

At this point, one can sum up the polarizations and obtain the decay width~\cite{Ivanov:2015woa, Tran:2024phq}
\begin{equation}
	\Gamma\left (V \to P\ell \bar{\nu}_{\ell} \right)
	= \frac{G_F^2}{(2\pi)^3}\frac{|V_{q_1q_2}|^2}{64m_1^3}
	\int\limits_{m^2_{\ell}}^{(m_1-m_2)^2}\!\!\!\! dq^2
	\int\limits_{s_1^-}^{s_1^+}\!\! ds_1
	\frac13 H_{\mu\nu} L^{\mu\nu},
	\label{eq:rate}
\end{equation}
where $s_1 =(p_P+p_{\ell})^2$ whose upper and lower limits are given by
\begin{equation}
	s_1^{\pm}=
	m_2^2+m_{\ell}^2-\frac{1}{2q^2}
	\left[(q^2-m_1^2+m_2^2)(q^2+m_{\ell}^2)
	\mp\lambda^{1/2}(q^2,m_1^2,m_2^2)\lambda^{1/2}(q^2,m_{\ell}^2,0)\right]
\end{equation}
with $\lambda(x,y,z) \equiv x^2+y^2+z^2-2(xy+yz+zx)$ being 
the K{\"a}ll{\'e}n function.  

However, as our objective in this paper is to explore the polarization observables in these decay, we will skip the last step in Eq.~(\ref{eq:rate}) and use the helicity amplitude technique to obtain a general $V\to P\ell\bar{\nu}_\ell$ twofold differential decay distribution in terms of $q^2$ and a polar angle $\theta$. The angle $\theta$ is defined in the $W^\ast$ rest frame as the angle between the momentum of the final charged lepton and the direction opposite to the daughter meson's momentum. One has 
\begin{equation}
\frac{d^2\Gamma}{dq^2 d\cos\theta} = 
\frac{|{\bf p_2}|}{(2\pi)^3 32m_1^2}\left(1-\frac{m_\ell^2}{q^2}\right)
\cdot\sum\limits_{\rm pol}|\mathcal{M}|^2
=\frac{G^2_F |V_{q_1q_2}|^2 }{(2\pi)^3}
\frac{|{\bf p_2}|}{192 m_1^2}\left(1-\frac{m_\ell^2}{q^2}\right)
H^{\mu\nu} L_{\mu\nu},
\label{eq:2-fold-dis}
\end{equation}
where  $|{\bf p_2}|=\lambda^{1/2}(m_1^2,m_2^2,q^2)/2m_1 $
is the momentum of the daughter meson in the $V$ rest frame. 

The Lorentz contraction $H^{\mu\nu} L_{\mu\nu}$ can be  evaluated using helicity amplitudes (see, e.g.,~\cite{Korner:1989ve, Faessler:2002ut, Ivanov:2019nqd}). One first defines an orthonormal and complete helicity basis
$\epsilon^\mu(\lambda_W)$ with three spin-1 components  orthogonal to
the momentum transfer $q^\mu$, i.e., $\epsilon^\mu(\lambda_W) q_\mu=0$, for 
$\lambda_W=\pm,0$, and one spin-0 (time) component $\lambda_W=t$ with
$\epsilon^\mu(t)= q^\mu/\sqrt{q^2}$. The orthonormality and completeness relations read 
\begin{eqnarray}
\epsilon^\ast_\mu(m)\epsilon^\mu(n) &=& g_{mn} 
\quad (\text{orthonormality}),
\\
\epsilon_\mu(m)\epsilon^{\ast}_{\nu}(n)g_{mn} &=& g_{\mu\nu} \quad\,\, (\text{completeness}),
\label{eq:orth-compl}
\end{eqnarray}
with $m,n=t,\pm,0$ and $g_{mn}={\rm diag}(+,-,-,-)$.
One then rewrites the tensor contraction using the completeness relation:
\begin{eqnarray}
L^{\mu\nu}H_{\mu\nu} &=& 
L_{\mu'\nu'}\epsilon^{\mu'}(m)\epsilon^{\ast\mu}(m')g_{mm'}
\epsilon^{\ast \nu'}(n)\epsilon^{\nu}(n')g_{nn'}H_{\mu\nu}
\nn
&=& L(m,n) g_{mm'} g_{nn'} H(m',n'),
\label{eq:contraction}
\end{eqnarray}
where $H(m,n)$ and $L(m,n)$ are the hadronic and leptonic tensors in the helicity-component space. One has
\begin{equation}
L(m,n) = \epsilon^\mu(m)\epsilon^{\ast \nu}(n)
L_{\mu\nu},
\qquad
H(m,n) = \epsilon^{\ast\mu}(m)\epsilon^\nu(n)H_{\mu\nu}.
\label{eq:hel_tensors}
\end{equation}
The point is that the two tensors can now be evaluated in two different
Lorentz frames. The hadronic tensor $H(m,n)$ will be evaluated in the parent $V$ rest frame while the leptonic tensor $L(m,n)$ {\textemdash} in the $W^\ast$ rest frame.

In the $V$ rest frame, the momenta and polarization vectors $\epsilon(\lambda_W)$ and $\epsilon_1(\lambda_V)$ can be written as
\begin{equation}
\begin{array}{lll}
	p^\mu_1 = (m_1,0,0,0),\quad & \quad
	\epsilon^\mu(t)\,\, =
	\frac{1}{\sqrt{q^2}}(q_0,0,0,|{\bf p_2}|),\quad
	& \quad 
	\\
	p^\mu_2 = (E_2,0,0,-|{\bf p_2}|), \quad &\quad
	\epsilon^\mu(\pm) = 
	\frac{1}{\sqrt{2}}(0,\mp 1,-i,0),\quad
	& \quad \epsilon^\mu_1(\pm) = 
	\frac{1}{\sqrt{2}}(0,\pm 1,-i,0),
	\\
	q^\mu   = (q_0,0,0,+|{\bf p_2}|), \quad &\quad
	\epsilon^\mu(0)\,\, =
	\frac{1}{\sqrt{q^2}}(|{\bf p_2}|,0,0,q_0),\quad &\quad \epsilon^\mu_1(0)\,\, = 
	(0,0,0,-1),\\
\end{array}
\label{eq:Dframe}
\end{equation}
where $E_2 = (m_1^2+m_2^2-q^2)/2 m_1$ and $q_0=(m_1^2-m_2^2+q^2)/2 m_1$.

The hadronic tensor $H(m,n)$ can now be written in terms of helicity amplitudes as:
\begin{eqnarray}
H(m,n) &=&  
\epsilon^{\ast \mu}(m) \epsilon^{ \nu}(n)H_{\mu\nu}
=
\epsilon^{\ast \mu}(m) \epsilon^{ \nu}(n) 
T^{VP}_{\mu\alpha}
\epsilon_1^{\alpha}(r)\epsilon_1^{\ast\beta}(s)\delta_{rs}
T^{VP\,\dagger}_{\beta\nu}
\nn
&=&
\epsilon^{\ast \mu}(m)\epsilon_1^{\alpha}(r)
T^{VP}_{\mu\alpha} \cdot
\Big[\epsilon^{\ast \nu}(n)\epsilon_1^{\beta}(s)T^{VP}_{\nu\beta}
\Big]^\dagger\delta_{rs}
\equiv 
H_{mr} H^{\dagger}_{nr},
\label{eq:hel_vv_def}
\end{eqnarray}
where the non-zero helicity amplitudes read
\begin{eqnarray}
H_{t0} &=&
\epsilon^{\ast \mu}(t)\epsilon_1^{ \alpha}(0)T^{VP}_{\mu\alpha}=
\frac{-|{\bf p_2}|}{(m_1+m_2)\sqrt{q^2}}
\left[(P\cdot q)(A_0+A_+)+q^2 A_-\right],\nn
H_{\pm\mp} &=&
\epsilon^{\ast \mu}(\pm)\epsilon_1^{\alpha}(\mp)T^{VP}_{\mu\alpha}=
\frac{(P\cdot q) A_0\mp 2m_1|{\bf p_2}| V}{m_1+m_2},\\
H_{00} &=&
\epsilon^{\ast \mu}(0)\epsilon_1^{\alpha}(0)T^{VP}_{\mu\alpha}= 
\frac{-(P\cdot q)(m_1^2 - m_2^2 + q^2) A_0 - 4m_1^2|{\bf p_2}|^2 A_+}{2m_1(m_1+m_2)\sqrt{q^2}}.\nonumber
\label{eq:hel_vv}
\end{eqnarray}

The leptonic tensor $L(m,n)$ is evaluated in the $W^\ast$ rest frame, where the charged lepton and the neutrino are back-to-back. It was calculated in great details in our previous study~\cite{Ivanov:2019nqd}. The result reads
\begin{eqnarray}
\label{lt2}
\lefteqn{(2q^2v)^{-1} L(m,n)(\theta)=}\\
&=&
\left( \begin{array}{cccc} 
	0 & 0 & 0 & 0 \\
	0 & (1-\cos\theta)^2 &0 & 0 \\
	0 & 0 & 2\sin^2\theta 
	& 0 \\
	0 & 0 &0 & (1+\cos\theta)^2 \\
\end{array} \right)
+\delta_\ell 
\left( \begin{array}{cccc} 
	4 & 0 & 4 \cos\theta &
	0 \\
	0 & 
	2 \sin^2\theta & 
	& 0 \\
	4\cos\theta & 
	0 & 
	4\cos^2\theta & 
	0 \\
	0 & 0 & 
	0 & 
	2\sin^2\theta \nn
\end{array} \right),
\end{eqnarray}
where we have introduced the parameter $v\equiv 1- m_\ell^2/q^2$ and the helicity-flip factor $\delta_\ell\equiv m_\ell^2/2q^2$. Note that the matrix columns and rows are ordered in the sequence 
$(t,+,0,-)$.

Finally, one obtains the differential decay distribution over $q^2$ and $\cos\theta$ as follows:
\begin{eqnarray}
\frac{d\Gamma(V\to P \ell\bar{\nu}_\ell)}{dq^2d(\cos\theta)} &=&
\frac{G_F^2 |V_{q_1q_2}|^2 |{\bf p_2}| q^2 v^2}{96 (2\pi)^3 m_1^2}
\Big\{
(1-\cos\theta)^2 |H_{+-}|^2 + (1+\cos\theta)^2 |H_{-+}|^2 
+2\sin^2\theta |H_{00}|^2
\nn
&&+ 2\,\delta_\ell \left[ \sin^2\theta (|H_{+-}|^2+|H_{-+}|^2) 
+ 2|H_{t0}-\cos\theta H_{00}|^2 \right]
\Big\}\nn
&\equiv& \frac{G_F^2 |V_{q_1q_2}|^2 |{\bf p_2}| q^2 v^2}{36 (2\pi)^3 m_1^2} \, W(\theta).
\label{eq:distr2}
\end{eqnarray}
We have defined $W(\theta)$ as the polar angular distribution. The twofold differential decay distribution~(\ref{eq:distr2}) allows us to define all physical observables in a systematic way.

\section{Physical observables and their predicted values}
\label{sec:result}
In this section, we demonstrate the technique one can use to extract information from the twofold angular decay distribution in Eq.~(\ref{eq:distr2}) in a systematic manner. This can be done by defining appropriate physical quantities that can be directly measured in experiments. We then provide predictions for their numerical values using the hadronic form factors obtained in the CCQM as already mentioned in Sec.~\ref{sec:formalism}. Note that the theoretical results obtained in this paper, such as Eq.~(\ref{eq:distr2}), are general and independent of form factors. One can use form factors from other approaches to derive numerical predictions. 

\begin{figure}[htbp]
	\centering
	\begin{tabular}{lr}
		\includegraphics[width=0.45\textwidth]{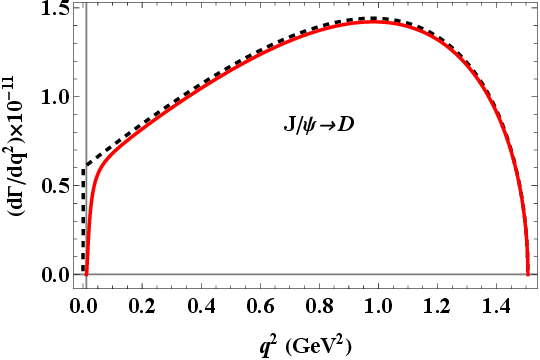} &
		\includegraphics[width=0.45\textwidth]{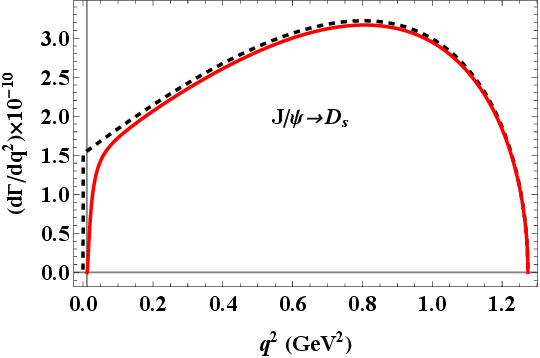}
	\end{tabular}
	\caption{\emph{$J/\psi\to D_{(s)}$ transition:} $q^2$ dependence of the decay rate $d\Gamma/dq^2$ for $e^-$ mode (dashed) and $\mu^-$ mode (solid).}
	\label{fig:difJpsi}
\end{figure}

\begin{figure}[htbp]
	\centering
	\begin{tabular}{lr}
		\includegraphics[width=0.45\textwidth]{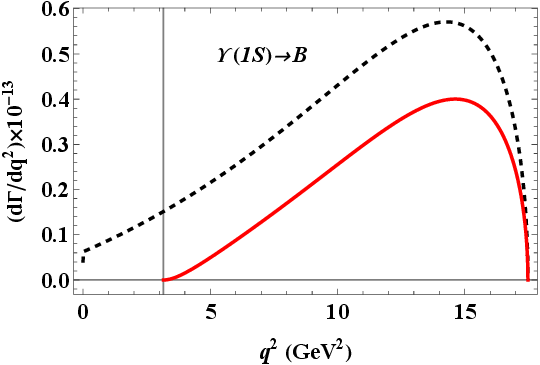} &
		\includegraphics[width=0.45\textwidth]{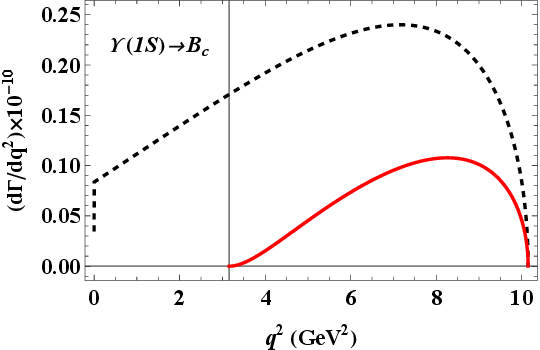}
	\end{tabular}
	\caption{\emph{$\Upsilon(1S)\to B_{(c)}$ transition:} $q^2$ dependence of the decay rate $d\Gamma/dq^2$ for $e^-$ mode (dashed) and $\tau^-$ mode (solid).}
	\label{fig:difUps}
\end{figure}

First, by integrating Eq.~(\ref{eq:distr2}) over $\cos\theta$ one obtains the $q^2$ decay distribution
\begin{equation}
	\frac{d\Gamma(V\to P \ell\bar{\nu}_\ell)}{dq^2} =
	\frac{G_F^2 |V_{q_1q_2}|^2 |{\bf p_2}| q^2 v^2}{36 (2\pi)^3 m_1^2}
	\Big[
	(1+\delta_\ell) (|H_{+-}|^2 + |H_{-+}|^2 
	+|H_{00}|^2) + 3\delta_\ell |H_{t0}|^2
	\Big].
	\label{eq:distr1}
\end{equation}
In Figures~\ref{fig:difJpsi} and \ref{fig:difUps} we show the decay distribution over $q^2$ for the decays $J/\psi\to D_{(s)}\ell\bar{\nu}_\ell$ and $\Upsilon(1S)\to B_{(c)}\ell\bar{\nu}_\ell$, respectively. Note that the $\tau^-$ mode is kinematically available only for the latter. Note also that the difference between the light lepton modes, i.e., $e^-$ and $\mu^-$, is small. Therefore, in the case of $\Upsilon(1S)\to B_{(c)}$ transitions, we only plot two lines: one for the $e^-$ mode and the other for the $\tau^-$ mode. The lepton-mass effect is clearly visible in this case.

\begin{table}[htbp]
	\centering
	\caption{Branching fractions of $J/\psi$ semileptonic decays in the CCQM and other frameworks.}
	\renewcommand{\arraystretch}{1.2}
	\begin{tabular}{c c c c c c c c}
		\hline\hline
		Mode & Unit &This work & LQCD~\cite{Meng:2024nyo} & BS~\cite{Wang:2016dkd}& QCDSR~\cite{Wang:2007ys} & LFQM~\cite{Shen:2008zzb} \\
		\hline
		$J/\psi \to D e \bar{\nu}_e$ & $10^{-11}$ & $1.71\pm 0.26$ & $1.21\pm 0.11$ & $2.03^{+0.29}_{-0.25}$ & $0.73^{+0.43}_{-0.22}$ 
		& $5.1 \sim 5.7$ \\
		$J/\psi \to D \mu \bar{\nu}_{\mu}$ & $10^{-11}$ & $1.66\pm 0.25$ & $1.18\pm 0.11$& $1.98^{+0.28}_{-0.24}$  & $0.71^{+0.42}_{-0.22}$ 
		& $4.7 \sim 5.5$ \\
		$J/\psi \to D_s e \bar{\nu}_e$ & $10^{-10}$ & $3.30\pm 0.50$ & $1.90\pm 0.08$ & $3.67^{+0.52}_{-0.44}$ & $1.8^{+0.7}_{-0.5}$ 
		& $5.3 \sim 5.8$ \\
		$J/\psi \to D_s \mu \bar{\nu}_{\mu}$ & $10^{-10}$ & $3.18\pm 0.48$ & $1.84\pm 0.08$ & $3.54^{+0.50}_{-0.43}$ & $1.7^{+0.7}_{-0.5}$ 
		& $5.5 \sim 5.7$ \\		
		\hline\hline
	\end{tabular}
	\label{tab:BrJpsi}
\end{table}
\begin{table}[htbp]
	\centering
	\caption{Branching fractions of $\Upsilon(1S)$ semileptonic decays in the CCQM and other frameworks.}
	\label{tab:BrUps}
	\renewcommand{\arraystretch}{1.2}
	\begin{tabular}{cccccc}
		\hline\hline
		Mode  & Unit & This work  &  BS~\cite{Wang:2016dkd} & BSW~\cite{Dhir:2009rb}& NRQCD~\cite{Chang:2016gyw}\\
		\hline
		$\Upsilon(1S)\to B e \bar\nu_e$ & $10^{-13}$ & $5.96\pm 0.89$ & $7.83^{+1.40}_{-1.20}$ &  &\\
		$\Upsilon(1S)\to B \mu \bar\nu_\mu$ & $10^{-13}$ & $5.95\pm 0.89$ & $7.82^{+1.40}_{-1.20}$ & &\\
		$\Upsilon(1S)\to B \tau \bar\nu_\tau$ & $10^{-13}$ & $3.30\pm 0.50$ & $5.04^{+0.92}_{-0.79}$ & &\\
		$\Upsilon(1S)\to B_c e \bar\nu_e$ & $10^{-10}$ & $1.84\pm 0.28$ & $1.37^{+0.22}_{-0.19}$ & $1.70^{+0.03}_{-0.02}$ & $5.58^{+3.32\,\,+0.14\,\,+0.08}_{-1.54\,\,-0.12\,\,-0.18}$\\
		$\Upsilon(1S)\to B_c \mu \bar\nu_\mu$ & $10^{-10}$ & $1.83\pm 0.27$ & $1.37^{+0.22}_{-0.19}$ & $1.69^{+0.04}_{-0.02}$ & $5.58^{+3.32\,\,+0.14\,\,+0.08}_{-1.54\,\,-0.12\,\,-0.18}$\\
		$\Upsilon(1S)\to B_c \tau \bar\nu_\tau$ & $10^{-11}$ & $4.74\pm 0.71$ & $4.17^{+0.58}_{-0.52}$ & $2.90^{+0.05}_{-0.02}$ & $13.0^{+7.7\,\,+0.3\,\,+0.2}_{-3.5\,\,-0.3\,\,-0.4}$\\
		\hline\hline
	\end{tabular}
\end{table} 
The decay distribution in Eq.~(\ref{eq:distr1}) can be integrated over the whole $q^2$ range and devided by the total decay width of the parent $V$ meson to obtain the branching fraction:
\begin{equation}
	\mathcal{B}(V\to P\ell\bar{\nu}_\ell) =\frac{1}{\Gamma_{\textrm{tot}}(V)} \int_{m_\ell^2}^{(m_1-m_2)^2}dq^2 \frac{d\Gamma (V\to P\ell\bar{\nu}_\ell) }{dq^2}.
\end{equation}
In Tables~\ref{tab:BrJpsi} and~\ref{tab:BrUps} we present the branching fractions predicted by the CCQM as well as by other approaches. Note that we already provided these predictions in our previous studies~\cite{Ivanov:2015woa, Tran:2024phq}. However, we then calculated these values using Eq.~(\ref{eq:rate}). In this paper, we apply the helicity amplitude technique and obtain the same results, providing a strong validation of our theoretical calculations. These tables also include recent predictions that were not available at the time of publication of Refs.~\cite{Ivanov:2015woa, Tran:2024phq}. Our results agree very well with those obtained in the BS approach~\cite{Wang:2016dkd}. Recently, the first LQCD calculation for $J/\psi\to D_{(s)}\ell\bar{\nu}_\ell$ has been provided in Ref.~\cite{Meng:2024nyo}. Our branching fractions agree with this LQCD calculation within $2\sigma$ in the case of $J/\psi\to D$. In the case of $J/\psi\to D_s$, the discrepancy is approximately $3\sigma$. 

For a better understanding of polarization observables which will be defined below, we introduce a normalized angular distribution
$\widetilde W(\theta)$ through
\begin{equation}
	\widetilde W(\theta)=\frac{W(\theta)}
	{ {\cal H}_{\rm tot}}, \qquad  {\cal H}_{\rm tot} \equiv (1+\delta_\ell) (|H_{+-}|^2 + |H_{-+}|^2 
	+|H_{00}|^2) + 3\delta_\ell |H_{t0}|^2.
	\label{eq:normdis}
\end{equation}
One can check that $\int_{-1}^{+1}\widetilde W(\theta)d\cos\theta=1$. The normalized $\theta$ distribution is described by a 
tilted parabola of the form
\begin{equation}
	\widetilde W(\theta)=a+b\cos\theta+c\cos^{2}\theta.
	\label{eq:theta-distr}
\end{equation}
The linear coefficient $b$ in Eq.~(\ref{eq:theta-distr}) can be extracted by defining a 
forward-backward asymmetry as follows:
\begin{eqnarray}
	\mathcal{A}_{FB}(q^2) = 
	\frac{d\Gamma(F)-d\Gamma(B)}{d\Gamma(F)+d\Gamma(B)}
	&=&
	\frac{ \int_{0}^{1} d\!\cos\theta\, (d\Gamma/d\!\cos\theta)
		-\int_{-1}^{0} d\!\cos\theta\, (d\Gamma/d\!\cos\theta) }
	{ \int_{0}^{1} d\!\cos\theta\, (d\Gamma/d\!\cos\theta)
		+\int_{-1}^{0} d\!\cos\theta\, (d\Gamma/d\!\cos\theta)} 
	\nn
	&=& b =-\frac34 \frac{|H_{+-}|^2-|H_{-+}|^2+4\delta_\ell H_{t0}H_{00}}{{\cal H}_{\rm tot}}.
	\label{eq:fbAsym}
\end{eqnarray}
\begin{figure}[htbp]
	\centering
	\begin{tabular}{lr}
		\includegraphics[width=0.45\textwidth]{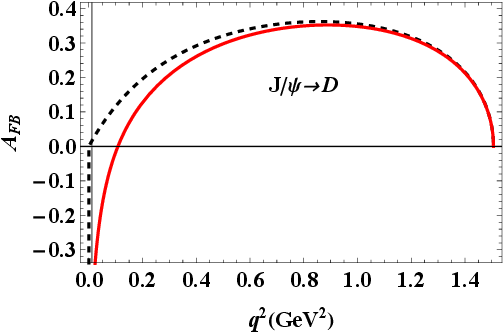} &
		\includegraphics[width=0.45\textwidth]{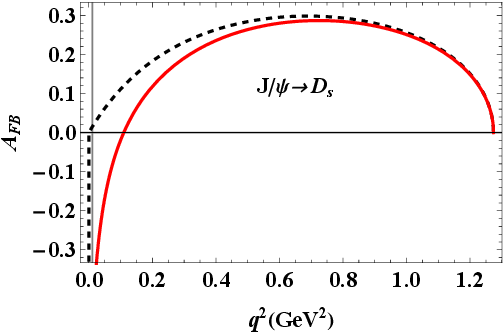}
	\end{tabular}
	\caption{\emph{$J/\psi\to D_{(s)}$ transition:} $q^2$ dependence of $\mathcal{A}_{FB}$ for $e^-$ mode (dashed) and $\mu^-$ mode (solid).}
	\label{fig:AFBJpsi}
\end{figure}
\begin{figure}[htbp]
	\centering
	\begin{tabular}{lr}		
		\includegraphics[width=0.45\textwidth]{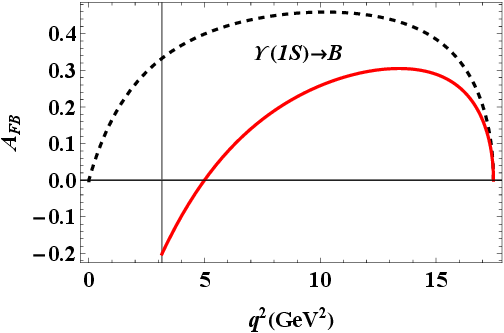} &
		\includegraphics[width=0.45\textwidth]{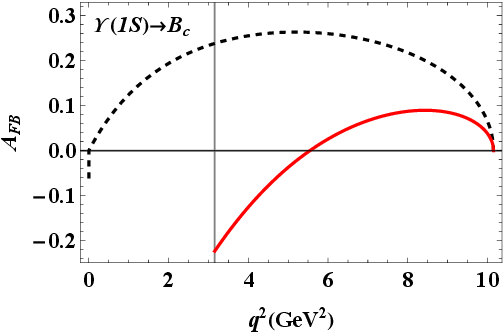}
	\end{tabular}
	\caption{\emph{$\Upsilon(1S)\to B_{(c)}$ transition:} $q^2$ dependence of $\mathcal{A}_{FB}$ for $e^-$ mode (dashed) and $\tau^-$ mode (solid).}
	\label{fig:AFBUps}
\end{figure}
The $q^2$ dependence of the forward-backward asymmetry is depicted in Figures~\ref{fig:AFBJpsi} and~\ref{fig:AFBUps}.

\begin{figure}[htbp]
	\centering
	\begin{tabular}{lr}
		\includegraphics[width=0.45\textwidth]{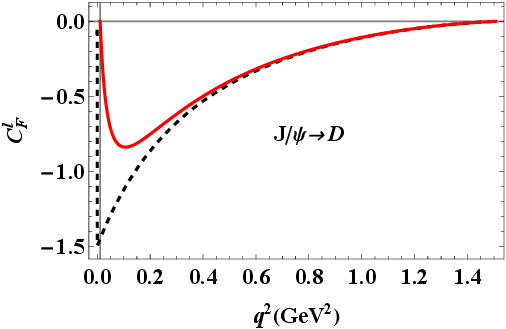} &
		\includegraphics[width=0.45\textwidth]{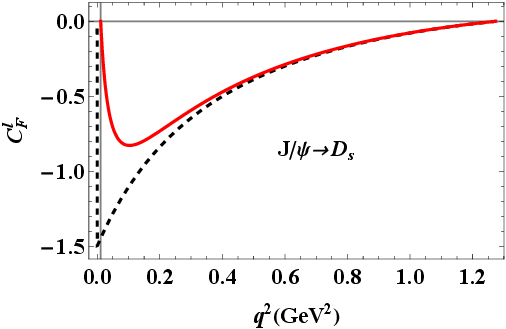}
	\end{tabular}
	\caption{\emph{$J/\psi\to D_{(s)}$ transition:} $q^2$ dependence of the lepton-side convexity parameter $C_F^\ell$ for $e^-$ mode (dashed) and $\mu^-$ mode (solid).}
	\label{fig:CFlJpsi}
\end{figure}

\begin{figure}[htbp]
	\centering
	\begin{tabular}{lr}	
		\includegraphics[width=0.45\textwidth]{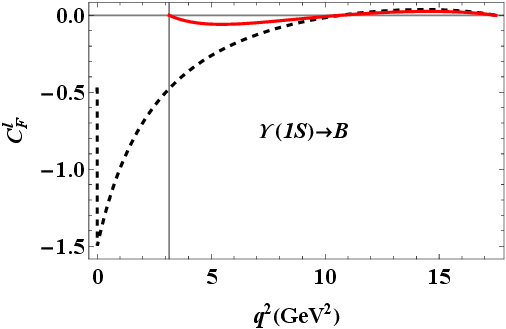} &
		\includegraphics[width=0.45\textwidth]{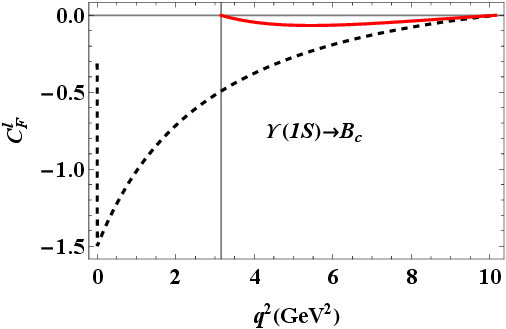}
	\end{tabular}
	\caption{\emph{$\Upsilon(1S)\to B_{(c)}$ transition:} $q^2$ dependence of the lepton-side convexity parameter $C_F^\ell$ for $e^-$ mode (dashed) and $\tau^-$ mode (solid).}
	\label{fig:CFlUps}
\end{figure}

The quadratic coefficient $c$ in Eq.~(\ref{eq:theta-distr}) is extracted by taking the second derivative of $\widetilde W(\theta)$. We therefore define a lepton-side convexity parameter by
\begin{equation}
	C_F^\ell(q^2) = \frac{d^{2}\widetilde W(\theta)}{d(\cos\theta)^{2}}
	= 2c
	= \frac34 (1-2\delta_\ell)
	\frac{ |H_{+-}|^2 + |H_{-+}|^2 
		-2|H_{00}|^2 }{ {\cal H}_{\rm tot} }.
	\label{eq:convex_lep}
\end{equation}
In Figures~\ref{fig:CFlJpsi} and~\ref{fig:CFlUps} we show the $q^2$ dependence of the lepton-side convexity for the decays $J/\psi\to D_{(s)}\ell\bar{\nu}_\ell$ and $\Upsilon(1S)\to B_{(c)}\ell\bar{\nu}_\ell$, respectively.

\begin{figure}[htbp]
	\centering
	\begin{tabular}{lr}
		\includegraphics[width=0.45\textwidth]{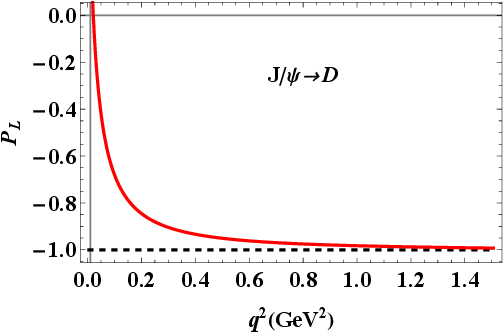} &
		\includegraphics[width=0.45\textwidth]{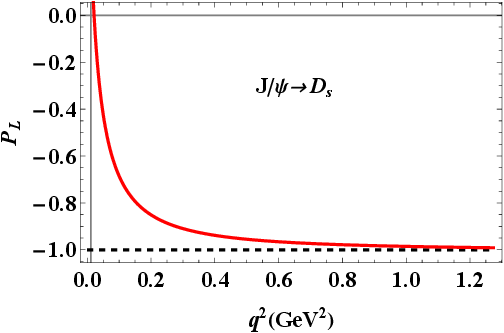}
	\end{tabular}
	\caption{\emph{$J/\psi\to D_{(s)}$ transition:} $q^2$ dependence of the longitudinal polarization of the final lepton $P_L^\ell$ for $e^-$ mode (dashed) and $\mu^-$ mode (solid).}
	\label{fig:PLJpsi}
\end{figure}

\begin{figure}[htbp]
	\centering
	\begin{tabular}{lr}		
		\includegraphics[width=0.45\textwidth]{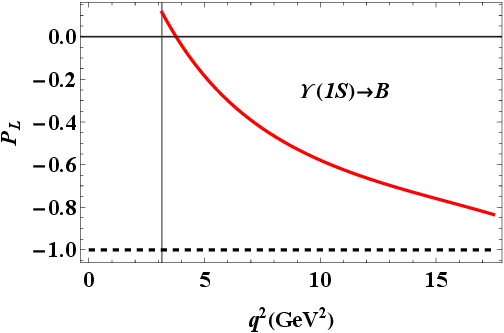} &
		\includegraphics[width=0.45\textwidth]{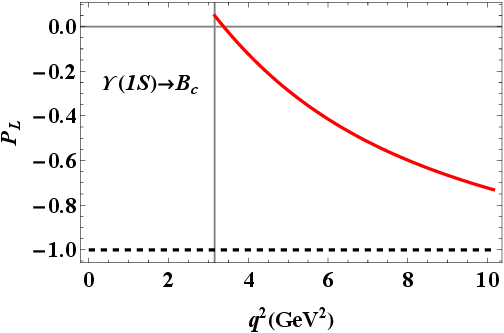}
	\end{tabular}
	\caption{\emph{$\Upsilon(1S)\to B_{(c)}$ transition:} $q^2$ dependence of the longitudinal polarization of the final lepton $P_L^\ell$ for $e^-$ mode (dashed) and $\tau^-$ mode (solid).}
	\label{fig:PLUps}
\end{figure}

Finally, we consider the longitudinal polarization $P_L^\ell(q^2)$ of the final lepton. The expression for this observable can be simply obtained from the difference between the helicity flip (${\cal H}_{\textrm{hf}}$) and non-flip (${\cal H}_{\textrm{nf}}$) structures in the total helicity $\cal{H}_{\textrm{tot}}$ given in Eq.~(\ref{eq:normdis}) as follows:
\begin{equation}
	P_L^\ell(q^2) = \frac{\delta_\ell {\cal H}_{\textrm{hf}}-{\cal H}_{\textrm{nf}}}{\delta_\ell {\cal H}_{\textrm{hf}}+{\cal H}_{\textrm{nf}}}
	= \frac{-(1-\delta_\ell)(|H_{+-}|^2 + |H_{-+}|^2 
		+|H_{00}|^2)+3\delta_\ell|H_{t0}|^2}{{\cal H}_{\rm tot}}.
\label{eq:PL}
\end{equation}
The $q^2$ dependence of the lepton longitudinal polarization is depicted in Figures~\ref{fig:PLJpsi} and~\ref{fig:PLUps}. Note that in the zero-lepton-mass limit ($m_\ell \to 0$) one has
\begin{equation}
	P_L^\ell(q^2) 
	= \frac{-(1-\delta_\ell)(|H_{+-}|^2 + |H_{-+}|^2 
		+|H_{00}|^2)+3\delta_\ell|H_{t0}|^2}{\phantom{-}(1+\delta_\ell) (|H_{+-}|^2 + |H_{-+}|^2 
		+|H_{00}|^2) + 3\delta_\ell |H_{t0}|^2}
	\xrightarrow[\delta_\ell\to 0]{} -1.
	\label{eq:PL}
\end{equation}
This limit is highly applicable to the case of electron, as can be seen from Figures~\ref{fig:PLJpsi} and~\ref{fig:PLUps}.

The $q^2$ dependence of the decay rate and polarization observables allows for probing the dynamics of the decays in great detail. However, at the first stage, one usually considers the integrated values for the observables over the entire $q^2$ range as they are less challenging from an experimental perspective. Note that when calculating the $q^{2}$ averages of the forward-backward asymmetry, convexity parameter, and lepton polarization, one 
has to multiply the numerator and denominator of Eqs.~(\ref{eq:fbAsym}),~(\ref{eq:PL}) and~(\ref{eq:convex_lep}) by the 
$q^{2}$-dependent piece of the phase-space factor 
$
C(q^2) = |\mathbf{p_2}| q^2 v^2
$
and integrate them separately.
For example, the $q^2$-mean forward-backward asymmetry can then be calculated 
according to
\begin{equation}
	\langle \mathcal{A}_{FB}\rangle = -\frac34 
	\frac{\int dq^{2} C(q^{2})\big(|H_{+-}|^2-|H_{-+}|^2+4\delta_\ell H_{t0}H_{00}\big)}
	{\int dq^{2} C(q^{2}){\cal H}_{\rm tot}}.
	\label{eq:FBint}
\end{equation}
Our predictions for the $q^2$ averages of polarization observables are summarized in Table~\ref{tab:obs-numerics}. We also list predictions obtained in Refs.~\cite{Chang:2016gyw, Sheng:2025duk} using the BSW model for comparison. Our predictions agree well with the BSW model in the case of the lepton polarization and convexity parameter. However, in the case of the forward-backward asymmetry, the results from two frameworks are somehow different.

\begin{table}[htbp] 
	\begin{center}
		\def\arraystretch{1.2}
		\vspace*{0.5cm}
		\begin{tabular}{c|ccc}
			\hline\hline	
			\quad Mode	\qquad	
			&\quad $<A_{FB}^\ell>$    \qquad   
			&\quad $<C_{F}^\ell>$\qquad 
			&\quad $<P_{L}^\ell>$ \qquad   
			\\
			\hline
			\qquad $J/\psi \to D e \bar{\nu}_e$\qquad  \quad   
			&\quad $ 0.30(3) $  \qquad  
			&\quad $ -0.31(3) $ \qquad 
			&\quad $  -1.0 $ \qquad  
			\\
			\qquad $J/\psi \to D \mu \bar{\nu}_{\mu}$ \qquad \quad   
			&\quad $ 0.28(3) $ \qquad  
			&\quad $ -0.26(3) $\qquad 
			&\quad  $  -0.94(9) $ \qquad  
			\\		
			\hline
			\qquad $J/\psi \to D_s e \bar{\nu}_e$\qquad  \quad   
			&\quad $0.24(2)$  \qquad  
			&\quad $-0.36(4)$ \qquad 
			&\quad $-1.0 $ \qquad  
			\\
			\qquad $J/\psi \to D_s \mu \bar{\nu}_{\mu}$ \qquad \quad   
			&\quad $0.22(2)$ \qquad  
			&\quad $ -0.30(3) $ \qquad 
			&\quad $ -0.93(9) $  \qquad  
			\\		
			\hline
			\qquad $\Upsilon(1S)\to B e \bar\nu_e$ \qquad  \quad   
			&\quad $ 0.40(4) $  \qquad  
			&\quad $ -0.076(8) $ \qquad 
			&\quad $  -1.0 $ \qquad  
			\\
			\qquad $\Upsilon(1S)\to B \mu \bar\nu_\mu$ \qquad \quad   
			&\quad $ 0.40(4) $ \qquad  
			&\quad $ -0.073(7) $\qquad 
			&\quad  $-1.0 $ \qquad  
			\\		
			\qquad $\Upsilon(1S)\to B \tau \bar\nu_\tau$\qquad  \quad   
			&\quad $0.25(3)$  \qquad  
			&\quad $0.0047(5)$ \qquad 
			&\quad $-0.65(7)$ \qquad  
			\\
			\hline
			\qquad $\Upsilon(1S)\to B_c e \bar\nu_e$ \qquad  \quad   
			&\quad $ 0.21(2) $  \qquad  
			&\quad $ -0.32(3)$ \qquad 
			&\quad $ -1.0 $ \qquad  
			\\
			&\quad $ \left[0.360^{+0.092}_{-0.077}\right]$  \qquad  
			&\quad $ \left[-0.374^{+0.013}_{-0.012}\right] $ \qquad 
			&\quad $ [-1.0 ]$ \qquad  
			\\	[2ex]
			\qquad $\Upsilon(1S)\to B_c \mu \bar\nu_\mu$ \qquad \quad   
			&\quad $ 0.21(2)$ \qquad  
			&\quad $ -0.31(3) $\qquad 
			&\quad  $  -1.0 $ \qquad  
			\\	
			&\quad $ \left[0.357^{+0.090}_{-0.075}\right]$ \qquad  
			&\quad $ \left[-0.361^{+0.013}_{-0.012}\right]$\qquad 
			&\quad  $ \left[-0.987^{+0.002}_{-0.002}\right] $ \qquad  
			\\		[2ex]
			\qquad $\Upsilon(1S)\to B_c \tau \bar\nu_\tau$\qquad  \quad   
			&\quad $ 0.052(5) $  \qquad  
			&\quad $-0.041(4) $ \qquad 
			&\quad $-0.53(5) $ \qquad  
			\\
			&\quad $\left[0.044^{+0.050}_{-0.040}\right]^{(\dagger)}$ \qquad  
			&\quad  \qquad 
			&\quad $ \left[-0.54^{+0.07}_{-0.04} \right]^{(\dagger)}$ \qquad 
			\\
			&\quad $ \left[0.069^{+0.050}_{-0.050}\right]$  \qquad  
			&\quad $ \left[-0.047^{+0.003}_{-0.003}\right]$ \qquad 
			&\quad $ \left[-0.537^{+0.023}_{-0.027} \right]$ \qquad  
			\\
			\hline\hline
		\end{tabular}
		\caption{$q^{2}$ averages of polarization observables. For comparison with 
			results using the BSW model~\cite{Sheng:2025duk,Chang:2016gyw}, we add in squared brackets the corresponding BSW values. Results form Ref.~\cite{Chang:2016gyw} are denoted with a dagger ($\dagger$). 
		}
		\label{tab:obs-numerics}
	\end{center}
\end{table}


\section{Summary}
\label{sec:sum}
We have studied in great detail the semileptonic decays
$J/\psi\to D_{(s)}\ell\bar{\nu}_\ell$ and $\Upsilon(1S)\to B_{(c)}\ell\bar{\nu}_\ell$ with special focus on the differential decay rate and polarization observables. We have applied the helicity amplitude technique to obtain a general twofold angular decay distribution for the $V\to P\ell\bar{\nu}_\ell$ decay, which allows for probing the dynamics of the last. A systematic approach to extract the information from the decay distribution has been introduced.
By using the invariant form factors of the $J/\psi\to D_{(s)}$ and $\Upsilon(1S)\to B_{(c)}$ hadronic transtions calculated in the Covariant Confined Quark Model, we have provided predictions for the polarization observables including the forward-backward asymmetry, lepton-side convexity parameter, and final lepton polarization. Most of our predictions are the first ones in the literature, except for the case of $\Upsilon(1S)\to B_c$, for which a few predictions using the BSW model are available. In general, our predictions for this case agree well with the BSW model within the provided uncertainties. Theoretical predictions given in this study are useful for future measurement of these decay channels, as well as for theoretical studies in other approaches.   

\section*{Acknowledgments}
This  research  is  funded  by  Vietnam National Foundation for Science and Technology Development (NAFOSTED) 
under grant number 103.01-2021.09. C.T.Tran thanks the organizers at Phenikaa University for the invitation and warm hospitality during the PIAS-2024 Workshop.

	
\end{document}